\documentclass[12pt]{article}

\usepackage[a4paper,left=30mm,right=20mm,top=20mm,bottom=20mm]{geometry}
\usepackage{graphics}
\usepackage{amsmath}
\usepackage{amssymb}
\usepackage{epsfig}
\usepackage{cite}
\usepackage{xcolor}
\usepackage{ctable}
\usepackage[unicode]{hyperref}
\graphicspath{{/}}

\title{Tree-level NLO corrections to inclusive $\psi'$ production in High Energy Factorization } 

\author{A.A.~Prokhorov$^{1}$, S.P.~Baranov$^{2}$, A.V.~Lipatov$^{3}$, M.A.~Malyshev$^{3,4}$, X.~Chen$^{5,6}$}

\begin{document}

\maketitle

\begin{center}

{\it $^{1}$Joint Institute for Nuclear Research, 141980 Dubna, Moscow region, Russia}\\
{\it $^{2}$P.N.~Lebedev Institute of Physics, 119991 Moscow, Russia}\\
{\it $^{3}$Skobeltsyn Institute of Nuclear Physics, Lomonosov Moscow State University, 119991 Moscow, Russia}\\
{\it $^{4}$Moscow Aviation Institute, 125993 Moscow, Russia}\\
{\it $^{5}$Institute of Modern Physics, Chinese Academy of Sciences, Lanzhou 730000, China}\\
{\it $^{6}$School of Nuclear Science and Technology, University of Chinese Academy of Sciences, Beijing 100049, China}

\end{center}

\vspace{0.5cm}

\begin{center}

{\bf Abstract }

\end{center}

We consider inclusive $\psi(2S)$ production in proton-proton collisions at collider
energies in the framework of non-relativistic QCD and high-energy factorization
beyond the low-order approximation.
We utilise a matching scheme proposed earlier to merge the leading order and
tree-level next-to-leading order production amplitudes and now extend it to low
transverse momenta and/or forward rapidities.
With the improved scheme, we examine the possibility to simultaneously describe all
unpolarized LHC data in the full kinematic range accessible to experimental
measurements and try different Transverse Momentum Dependent (TMD) gluon distributions in the proton.
A global fit to the data is carried out to extract the color octet long-distance
matrix elements for $\psi(2S)$ mesons. We show that taking the NLO corrections
into account leads to better description of the data.

\vspace{1cm}


\newpage

\section{Introduction}\indent

Understanding the data on the inclusive and associated
production of heavy quarks bound states --- heavy quarkonia --- is one of topical
subjects of modern high energy physics\cite{Quarkonia-Review-1, Quarkonia-Review-2,
Quarkonia-Review-3}.
Among the different approaches to the problem, the formalism of nonrelativistic QCD
(NRQCD)\cite{NRQCD-1, NRQCD-2,NRQCD-3} is commonly regarded as providing the most
rigorous working framework. For the sake of completeness one also should mention the
color singlet model (CSM) \cite{CSM-1, CSM-2, CSM-3} (actually, a predecessor of
NRQCD) and the color evaporation model (CEM)\cite{CEM-1, CEM-2, CEM-3, CEM-4}.
The NRQCD formalism exploits a hypothesis that the perturbative production of heavy quarks
in a hard parton-parton interaction factorises from their subsequent non-perturbative
evolution to a physical meson. The probabilities of these two sequential steps are
given, accordingly, by the partonic cross section $\sigma(Q\bar{Q})$ and the
long-distance matrix element (LDME).

In calculating the perturbative step, we employ the
$k_T$-factorization\cite{kt-factorization} (or high energy factorization\cite{HighEnergyFactorization}) approach where the gluon
distributions may obey Balitsky-Fadin-Kuraev-Lipatov (BFKL)\cite{BFKL}
or Ciafaloni-Catani-Fiorani-Marchesini (CCFM)\cite{CCFM} evolution equations
and are called Transverse Momentum Dependent (TMD) or unintegrated.
This approach can be considered as convenient alternative to explicit high order
pQCD calculations. It effectively includes
the leading-log part of higher order corrections (namely, the terms proportional to
$\alpha_s^n \ln^n s/\Lambda_{\rm QCD}^2 \sim \alpha_s^n \ln^n 1/x $ up to infinitely
large $n$) absorbing them into TMD gluon densities. These corrections correspond to
the emission of real gluons in the initial state and play the dominant role at high
energies $s$ (or, equivalently, at small $x$). A detailed description of the
$k_T$-factorization technique can be found, for example, in reviews\cite{TMD-Review,
TMD-Review2}.

The incorporation of the $k_T$-factorization approach with the NRQCD formalism
enables one to achieve an improved overall description of the collider data on heavy
quarkonia. In fact, the description obtained in the $k_T$-factorization already with
LO off-shell amplitudes for inclusive charmonia 
and bottomonia 
production (see\cite{Our-Charmonia-1, Our-Charmonia-2, Our-Charmonia-3,
Our-Bottomonia-1, Our-Bottomonia-2, Our-Bottomonia-3} and references therein)
is of not poorer quality than that seen in the collinear factorization at NLO
\cite{Charmonia-NRQCD-1, Charmonia-NRQCD-2, Charmonia-NRQCD-3, Charmonia-NRQCD-4,
Charmonia-NRQCD-5, Charmonia-NRQCD-6, Charmonia-NRQCD-7, Charmonia-NRQCD-8,
Bottomonia-NRQCD-1, Bottomonia-NRQCD-2, Bottomonia-NRQCD-3, Bottomonia-NRQCD-4,
Bottomonia-NRQCD-5, Bottomonia-NRQCD-6}, see also \cite{Quarkonia-Review-2}).
Employing the special depolarization model\cite{TransitionMechanism} based on
the classical multipole radiation theory made it possible to obtain consistent
predictions for $J/\psi$ and $\psi'$ polarization observables, thus providing
a solution to a long-standing "polarization puzzle".
Further progress has recently been made\cite{Our-Chic-NLO} in restoring the Heavy
Quark Spin Symmetry (HQSS) relations\cite{HQSSPuzzle-1, HQSSPuzzle-2} in the
$\chi_c$ production. Violations of these relations were present in our previous
studies\cite{UnEqualWaveFunctions, Our-Charmonia-1, Our-Charmonia-3}, but have
gone away after including the NLO contributions.

In the present study we turn to the inclusive production of $\psi^\prime$ mesons
at the LHC conditions. A recent analysis\cite{GlobalFit_Kniehl} performed in the
conventional (collinear) factorization scheme at NLO shows that a reasonable
agreement with the data (polarized and unpolarized) can only be achieved for a
limited $p_T$ region, $p_T(\psi') > 7$ TeV, while the theory strongly overestimates
the data at smaller $p_T(\psi')$. The same kind of discrepancy is also seen in the
$k_T$-factorization predictions at LO\cite{Our-Charmonia-1}  for definite sets of TMD gluon distributions. Our present goal is to
improve the theoretical calculations by implementing the NLO corrections.

A major difficulty in including higher order corrections in the $k_T$-factorization
is to avoid double counting,  i.e., the cases when the same gluon emission is
counted twice\cite{Our-Chic-NLO}: as part of the initial gluon evolution and as
part of the hard partonic subprocess. Therefore, one needs to take care to separate
the gluon emissions in the phase space and then to accurately merge the LO and NLO
amplitudes. We basically follow the procedure\cite{Our-Chic-NLO} which has already
been successfully used for inclusive $\chi_c$ production. Now, with some proper
modifications, we extend it to relatively low transverse momenta and/or forward
rapidities.
We wish to perform a global fit to all available unpolarized $\psi'$ data, leaving
the polarization analysis for a following dedicated study\footnote{The only known data
on $\psi'$ polarization are collected by the CMS\cite{CMS_polarization, CMS_polarization_2} at $\sqrt s=7,\, 13$~TeV  and
LHCb\cite{LHCb_polarization} Collaborations at $\sqrt s=7$~TeV. These data contain a small amount of points in comparison with unpolarized (measurements of polarization observables in different frames may not be considered as independent) and show large experimental uncertainties. So, they cannot affect
the global fit and, for the time being, can be ignored.}.
The data included in the fit are taken at $\sqrt s = 5.02$, $7$, $8$ and $13$~TeV and cover
the entire rapidity range\cite{dataset_ATLAS_5, dataset_ATLAS7_8, dataset_ATLAS13, dataset_CMS_5, dataset_CMS7,
dataset_CMS13, dataset_LHCb7_13, dataset_ALICE7, dataset_ALICE8, dataset_ALICE13}.
Such an extensive analysis of the $\psi^{\prime}$ production 
in the $k_T$-factorization approach with taking into account NLO
terms is performed for
the first time.

The paper is organized as follows. In Section 2 we briefly describe our theoretical
framework and the basic steps of our calculations. In Section 3 we present the
numerical results and discussions.
Section 4 sums up our conclusions.

\section{Theoretical framework} \indent

This section provides a brief review of the $k_T$-factorization formalism applied
to $\psi^{\prime}$ production, the essential calculation steps, and the fitting
procedure. In general, our method of calculations resembles the usual NLO$^*$ (tree-level NLO)
scheme, though with two notable exceptions. First, in the $k_T$-factorization
approach, the true LO contribution to the CO production mechanism is represented by
a $2\to 1$ off-shell gluon-gluon fusion subprocess, while in the collinear case the
perturbation series starts from $2\to 2$ subprocesses (see below). Accordingly, our
$2\to 2$ subprocesses are regarded as NLO corrections. Second, our calculations are
also made in the tree-level approximation, but for a different reasoning and in a
different way as compared to the usual NLO$^*$. The kinematic constraints which
reject virtual and infrared singularities arise from a double-counting-exclusion (DCE) condition, which separates the LO and NLO contributions
(see also\cite{MatchingProcedure-1,
MatchingProcedure-2}). For the sake of definiteness and clarity, we hereafter will
refer to this scheme as to NLO$^\dag$\cite{Our-Chic-NLO}.

\subsection{Basic formulas} \indent

The LO contributions to $\psi^{\prime}$ production are presented by several
$\mathcal{O}(\alpha_s^2)$ and $\mathcal{O}(\alpha_s^3)$ off-shell gluon-gluon fusion
subprocesses producing the $c\bar{c}$ pairs in the CO and CS states, respectively:
\begin{gather}
  g^*(k_1)+g^*(k_2)\to c\bar c\left[ \, ^1S_0^{[8]},\, ^3P_J^{[8]},\, ^3S_1^{[8]}\right] (p),\label{eq:LO1} \\
  g^*(k_1)+g^*(k_2)\to c\bar c\left[ \, ^3S_1^{[1]}\right](p) + g(p_g),
  \label{eq:LO2}
\end{gather}
\noindent
where the four-momenta of all particles are indicated in the parentheses and
$J = 0$, $1$, $2$.
It is important that the interacting gluons have non-zero transverse momenta,
non-zero virtualities $k_1^2 = - {\mathbf k}_{1T}^2 \neq 0$,
$\;k_2^2 = - {\mathbf k}_{2T}^2 \neq 0$, and longitudinal components in their
polarization vectors. The tree-level NLO corrections are represented by a
$2 \to 2$ color octet $\mathcal{O}(\alpha_s^3)$ subprocesses:
\begin{gather}
	g^*(k_1) + g^*(k_2)\to c\bar c\left[\, ^1S_0^{[8]},\, ^3P_J^{[8]},\, ^3S_1^{[8]}\right](p) + g(p_g).
	\label{eq:NLO}
\end{gather}
\noindent
If the produced $c\bar c$ pair forms a color singlet state ${^3S}_1^{[1]}$,
it can directly convert into $\psi^\prime$ meson with the probability given by the
radial wave function at the origin of the coordinate space, $|R_\psi(0)|^2$
(see\cite{CSM-1, CSM-2, CSM-3}). The 
latter can be extracted from the measured $\psi^\prime$ leptonic decay width or
calculated within the potential models\cite{PotentialModelCalcutations-1,
PotentialModelCalcutations-2}.
The CO states evolve into real $\psi^\prime$ mesons through non-perturbative QCD
transitions (via soft gluon emissions). According to the multipole radiation
scenario\cite{TransitionMechanism}, this step is described in terms of
chromo-electric dipole ($E1$) transitions\cite{ProjectionOperators-4, ProjectionOperators-5}, which dominate the multipole expansion.
In the case of $^3P_J^{[8]}$ states, a single $E1$ transition
$c\bar c \left[{^3P}_J^{[8]}\right] \to c\bar c \left[{^3S}_1^{[1]}\right] + g$
is needed to transform them into $\psi^\prime$ mesons, whereas the trasformation of
$^3S_1^{[8]}$ state is treated as two sequential transitions:
$c\bar c\left[{^3S}_1^{[8]}\right] \to c\bar c \left[{^3P}_J^{[8]}\right] + g$ and
$c\bar c \left[{^3P}_J^{[8]}\right] \to c\bar c \left[{^3S}_1^{[1]}\right] + g$,
proceeding via either of three intermediate CO states: $J = 0, 1, 2$.
It is essential in our model that the gluons are not infinitely soft, but carry
some finite energy $E_g \sim \Lambda_{\rm QCD}$, that ensures escaping the domain
of confinement (see also\cite{TransitionMechanism}). The intermediate color octet
state must then be heavier than the final state quarkonium. The energy required to
overcome the confinement is a model parameter which brings additional uncertainty
to numerical calculations (see discussion below).

The expressions for the off-shell production amplitudes (\ref{eq:LO1}) --- (\ref{eq:NLO})
can be found\cite{Our-Charmonia-1, Our-Chic-NLO},
where their gauge invariance has been specially tested.
The latter
is achieved by using effective vertices (see\cite{EffectiveVertices1, EffectiveVertices2} and
references therein),
that ensure the exact gauge invariance even
with the off-shell initial gluons.
The probabilities
to form definite bound states are given by the corresponding LDME's,
$\langle\mathcal{O}^{\psi^\prime}[n]\rangle$, where $n$ is the proper Fock state.
The details of calculations are described\cite{Our-Charmonia-1, Our-Chic-NLO}.
Here we note that the proposed approach results in a good description
of the available LHC data on charmonia and bottomonia polarizations (see\cite{Our-Charmonia-1, Our-Charmonia-2, Our-Charmonia-3, Our-Bottomonia-1, Our-Bottomonia-2, Our-Bottomonia-3}).

According to the $k_T$-factorization
prescription\cite{kt-factorization, HighEnergyFactorization},
the cross section of a considered processes is calculated as a convolution of an
off-shell production amplitude $\bar {|{\cal A}|^2}$ and TMD gluon densities in
the proton, $f_g(x, {\mathbf k}_T^2, \mu^2)$.
Thus, the cross sections for the $2 \to 1$ and $2 \to 2$ subprocesses   (\ref{eq:LO1}) --- (\ref{eq:NLO}) can be written as:
\begin{gather}
  \sigma_{2 \to 1} = \int{2\pi \over x_1\,x_2\,s\,F}\; f_g(x_1, {\mathbf k}_{1T}^2, \mu^2) f_g(x_2, {\mathbf k}_{2T}^2, \mu^2) \overline{|{\cal  A}_{2\to1}|^2} \, d{\mathbf k}_{1T}^2 d{\mathbf k}_{2T}^2 dy {d\phi_1 \over 2\pi} {d\phi_2 \over 2\pi}, \\
  \sigma_{2 \to 2} = \int{1\over 8\pi\,x_1\,x_2\,s\,F}\;f_g(x_1, {\mathbf k}_{1T}^2, \mu^2) f_g(x_2, {\mathbf k}_{2T}^2, \mu^2) \overline{|{\cal  A}_{2\to2}|^2} \times \nonumber \\
  \times d{\mathbf p}_{T}^2 d{\mathbf k}_{1T}^2 d{\mathbf k}_{2T}^2 dy dy_g {d\phi_1 \over 2\pi} {d\phi_2 \over 2\pi},
\end{gather}
\noindent
where $\phi_1$ and $\phi_2$ are the azimuthal angles of the initial off-shell gluons,
$p_T$ and $y$ are the transverse momentum and rapidity of the produced $\psi'$ meson,
$y_g$ is the rapidity of the outgoing gluon, $\sqrt s$ is the $pp$ center-of-mass
energy, $\mu$ is the factorization scale and
$F = 2\lambda^{1/2}(\hat s, k_1^2, k_2^2)$ is the
flux factor\footnote{In the case of $2 \to 2$ processes,
this formula approximates standard expression $F \sim x_1 x_2 s$, see
discussion\cite{Flux-2} for more details.}, where $\hat s = (k_1 + k_2)^2$\cite{Flux-1}.
The necessary merging procedure for $2 \to 1$ and $2 \to 2$ contributions is discussed below.

Our calculations refer to two sets of TMD gluon densities, namely, A0\cite{A0} and
JH'2013 set 2\cite{JH2013}.
All these gluon distributions are obtained from a numerical solution of the CCFM
equation (at the leading logarithmic approximation, LLA\footnote{Unfortunately, the next-to-leading logarithmic corrections
to the CCFM equation are yet not known. However, it can be argued\cite{CASCADE2}
that the CCFM evolution at the LLA leads to reasonable QCD predictions.}) and are widely used in
phenomenological applications (see, for example,\cite{Motyka-photon, LMJ-PP,
LM-Higgs, LLM-FL, LLM-photon} and references therein). The parameters of (rather
empirical) input distributions used in the JH'2013 set 2 and A0 sets were derived from
a fit of the proton structure functions $F_2(x, Q^2)$ and $F_2^c(x, Q^2)$ to HERA
data at small $x$. All these TMD gluon densities are available from a commonly
known package \textsc{tmdlib} \cite{TMDLib2}, which is a C$++$ library providing
a framework and an interface to many different
parametrizations.
They are implemented also into the Monte-Carlo event generator \textsc{pegasus}\cite{PEGASUS}.

\subsection{Merging the LO and NLO contributions} \indent

In this section we discuss the procedure for merging the $2 \to 1$ (LO) and
$2 \to 2$ (NLO) contributions, which is necessary to avoid double counting in the
$k_T$-factorization calculations.
We follow the scheme which was first proposed and successfully applied
to $\chi_{cJ}$ production (see  \cite{Our-Chic-NLO} for more details).
The key idea of this method is to introduce double-counting-exclusion (DCE) cut for
$2\to 2$ subprocess: $|{\mathbf p}_{gT}| > \max(|{\mathbf k}_{1T}|,|{\mathbf k}_{2T}|)$.
It excludes the gluons already generated by the CCFM evolution in $2 \to 1$ and ensures that the hardest gluon emission in $2 \to 2$ events comes from the
hard matrix elements.

\begin{figure}
\begin{center}
{\includegraphics[width=.5\textwidth]{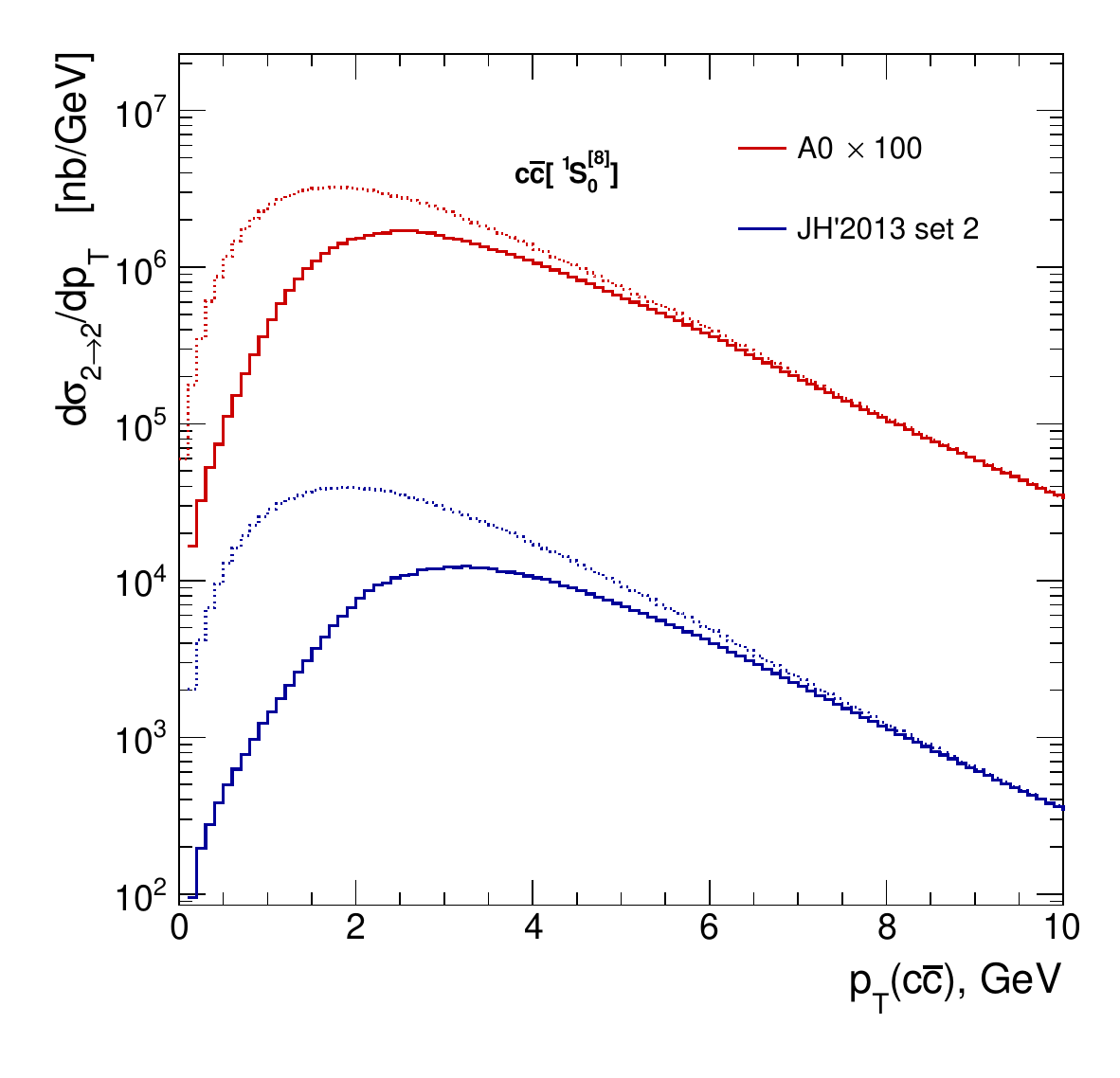}}\hfill
{\includegraphics[width=.5\textwidth]{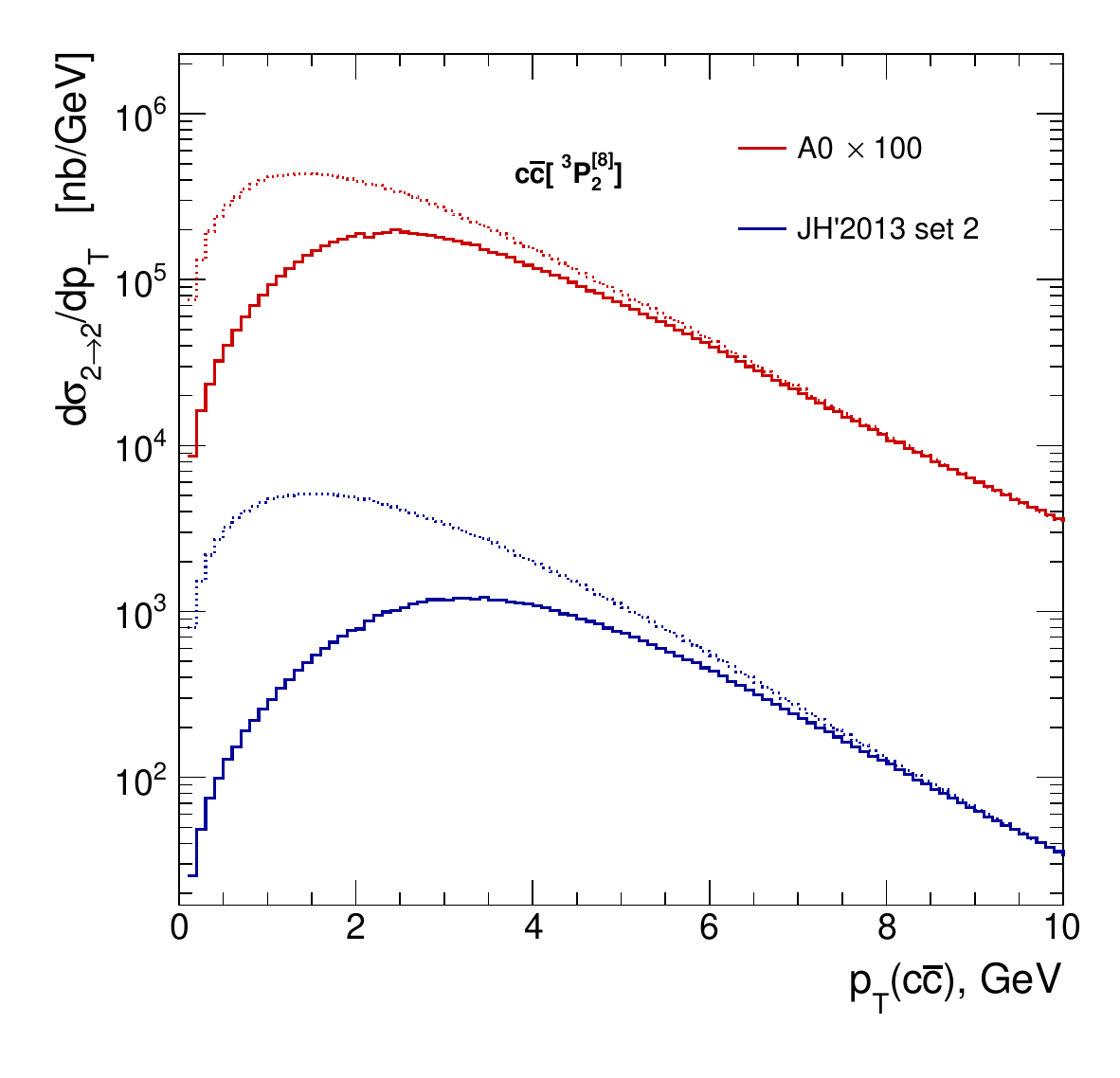}}\hfill
\caption{Comparison of $d\sigma_{2\rightarrow 2}/dp_{T}$ contributions for color octer $c\bar{c}[^1S^{[8]}_0]$ (left panel) and $c\bar{c}[^3P^{[8]}_2]$ (right panel) with standard (dashed) and enhanced DCE cuts (solid) in the forward rapidity region $2 < y < 4.5$ at $\sqrt{s} = 13$ TeV.}
\label{fig:DCE_comparison}
 \end{center}
\end{figure}

The method works well in the central region, but there exist specific kinematic
conditions where such DCE cut may not be sufficient.
Let us consider the forward production at low $p_{T}$ (say, less than $5$ or
$10$~GeV). Here, one of the incoming gluons $k_2$ (the one moving in the direction
opposite to the produced quarkonium) carries significantly lower momentum fraction
than the other gluon $k_1$ ($x_2 \ll x_1$, up to several orders of magnitude).
The evolution of $k_1$ gluon differs from that of the gluon $k_2$.
In particular, there is a smearing of the angular ordering\footnote{In case of CCFM evolution the gluon emissions obey the angular ordering condition arising from color coherence effect}, that affects the
transverse momenta of the gluons belonging to the $k_2$ evolution branch.
The primordial gluon transverse momentum\footnote{We mean, the momentum of the
gluon at the starting scale $\mu_0$, before the first step of the evolution.},
${\mathbf k}^{\rm init}_{T}$, becomes more important and should be taken into
account.
To safely avoid double counting under these conditions, we should strengthen the
DCE cut as $|{\mathbf p}_{gT}| > \max (|{\mathbf k}_{1T}|,|{\mathbf k}_{2T}|+|\overline{{\mathbf k}^{\rm init}_{2T}}|)$.
This shift is based on a simple kinematic relation for gluon emissions
${\mathbf k}^{\rm init}_{2T} =  \sum{\mathbf p}^{\rm emission}_{gT_i}+ {\mathbf k}_{2T}$
together with an estimation of the transverse momentum of the hardest gluon in the
evolution
$|{\mathbf p}^{\rm emission}_{gT}| < |{\mathbf k}_{2T}| + |{\mathbf k}^{\rm init}_{2T}|$.
The parameter $|\overline{{\mathbf k}^{\rm init}_{2T}}|$ represents the average
transverse momentum of the primordial gluon and can be extracted from the TMD gluon
distributions at the starting scale $\mu_0$:
$|\overline{{\mathbf k}^{\rm init}_{T}}| = 1.12$~GeV and $1.77$~GeV for A0 and
JH'2013 set 2 gluons, respectively.
Such DCE cut efficiently excludes $2 \to 2$ events where the gluon emitted at the
last step of the evolution has smaller transverse momentum than the primordial gluon.
In other words, gluons acquiring in the evolution larger $p_T$ than the gluons
coming from the hard subprocess are rejected by the merging scheme. Such gluons
are already taken into account in the LO part of the calculations. This way, we
safely avoid double counting in the low $p_T$ region.

The influence of the strenghtened DCE cut on the NLO$^{\dag}$ contributions from $^1S^{[8]}_0$
and $^3P^{[8]}_2$ intermediate states is shown in Fig.~\ref{fig:DCE_comparison}.
The calculations were done for relatively low $p_T$ in the forward rapidity region,
$ 2 < y^{\psi} < 4.5$.
The modified (strenghtened) scheme converges to the non-modified scheme at
$p_{T} > 6$~GeV.
We thus conclude that the modified DCE cut provides a somewhat more accurate
treatment at relatively low $p_T$ and/or forward rapidities and smoothly goes
to the previous scenario\cite{Our-Chic-NLO} at larger transverse momenta.
Note that using of strenghtened DCE cut is only
improvement of our present calculations. 


 \subsection{Numerical parameters} \indent

Following\cite{PDG}, we set the meson mass to $m(\psi^{\prime}) = 3.686$~GeV
and branching fraction $Br(\psi^{\prime} \to \mu^+\mu^-) = 0.8$\%.
We use one-loop expression for the QCD coupling $\alpha_s$ with $n_f = 4$ quark
flavours and $\Lambda_{\rm QCD} = 250$~MeV for A0 gluon and two-loop expression
with $n_f = 4$ and $\Lambda_{\rm QCD} = 200$~MeV for JH'2013 set 2 gluon
distribution\cite{A0, JH2013}.
Our default choice for the renormalization scale $\mu_R$ is the transverse mass $m_T^\psi$ of
the produced meson. The factorization
scale $\mu_F$ was set to $\mu^2_F = \hat s + {\mathbf Q}_T^2$, where ${\mathbf Q}_T$
is the net transverse momentum of the
initial off-shell gluon pair. Such a choice of $\mu_F$ is specific for CCFM
evolution (see\cite{CCFM} for more details).
However, in accordance with the proposed merging LO $+$ NLO
scheme\cite{Our-Chic-NLO}, we shift the scale as
$\mu_F \rightarrow \mu_F=m^{\psi}_T$ for 2 $\to$ 2 CO
subprocesses. 
The CS LDME is determined from the meson decay width\cite{Charmonia_width}:
$\langle\mathcal{O}^{\psi^{\prime}}[^3S^{[1]}_1]\rangle = 0.73$ GeV$^3$.

\subsection{LDME's fitting procedure}\indent

The values of LDME's for different CO intermediate states are free parameters not
predictable by the theory and have to be found from the experimental data.
As it was said, we perform a global fit which includes all of the available
unpolarized LHC data on inclusive $\psi^{\prime}$ production.
There are only $3$ such free parameters, namely, $\langle\mathcal{O}^{\psi^{\prime}}[^1S^{[8]}_{0}]\rangle, \langle\mathcal{O}^{\psi^{\prime}}[^3P^{[8]}_{0}]\rangle $
and $\langle\mathcal{O}^{\psi^{\prime}}[^3S^{[8]}_{1}]\rangle $, while the other
LDME's can be obtained from HQSS relations:
\begin{gather}
\langle\mathcal{O}^{\psi'}[^3P^{[8]}_{J}]\rangle\; =\; (2J+1)\,\langle\mathcal{O}^{\psi'}[^3P^{[8]}_{0}]\rangle\;
\label{eq:HQSS}
\end{gather}
\noindent
where $J = 0, 1, 2$. According to the NRQCD formalism, the $\psi^{\prime}$
production cross section reads:
\begin{gather}
\frac{d\sigma^{\psi^{\prime}}}{dp_{T}} = \frac{d\sigma(c\bar{c}[^3S^{[1]}_1])}{dp_{T}}\langle\mathcal{O}^{\psi^{\prime}}[^3S^{[1]}_{1}]\rangle + \frac{d\sigma(c\bar{c}[^1S^{[8]}_0])}{dp_{T}}\langle\mathcal{O}^{\psi^{\prime}}[^1S^{[8]}_{0}]\rangle + \nonumber \\
 + \sum_{J}(2J+1)\frac{d\sigma(c\bar{c}[^3P^{[8]}_J])}{dp_{T}}\langle\mathcal{O}^{\psi^{\prime}}[^3P^{[8]}_{0}]\rangle +\frac{d\sigma(c\bar{c}[^3S^{[8]}_1])}{dp_{T}}\langle\mathcal{O}^{\psi^{\prime}}[^3S^{[8]}_{1}]\rangle.
\end{gather}
\noindent
The values of LDME's are determined by minimizing the $\chi^2$ function running
over all datapoints. The uncertainties are calculated from the error matrix.
The fitted values of LDME's show strong negative correlation between
$^1S^{[8]}_{0}$ and $^3P^{[8]}_{J}$ contributions for both considered TMD densities.
This makes it impossible to extract these parameters with acceptable accuracy. Therefore, we replace the individual parameters with a linear combination:
\begin{gather}
  \langle\mathcal{O}_{l.c.}\rangle = \langle\mathcal{O}[^3P^{[8]}_{0}]\rangle + r \times \langle\mathcal{O}[^1S^{[8]}_{0}]\rangle,
\end{gather}
 \noindent
where $r$ is defined as
\begin{gather}
r = \frac{d\sigma(c\bar{c}[^1S^{[8]}_0])/dp_{T}}{\sum (2J+1)d\sigma(c\bar{c}[^3P^{[8]}_J])/dp_{T}}.
\end{gather}
\noindent
The coefficient $r$ was calculated for every datapoint.
Its values were averaged over the whole rapidity range and, separately, over the
central and forward rapidity regions\footnote{The differences in the values of $r$
in the central and forward regions reflect only differences in the normalization
in the different regions of the phase space, while the shapes of the transverse
momentum distributions remain the same. This does not affect the obtained results
but can be useful for upcoming studies.}.
We present them in Table~\ref{tab:r_ratio} for both A0 and JH'2013 set 2 gluons.

The results presented in the next section are obtained with only two free parameters,
$\langle\mathcal{O}_{l.c.}\rangle$ and $\langle\mathcal{O}[^3S^{[8]}_{1}]\rangle$,
which can be extracted with a good accuracy.

\begin{table}[ht]
	\centering
	\begin{tabular}[t]{lccc}
		\hline
		TMD &  \ central  & \ forward & \ combined \\
		\hline
		\hline
		A0  & $ 0.51 \pm 0.04 $ & $ 0.64 \pm 0.07$ & $ 0.54 \pm 0.07$  \\
        JH'2013 set 2  & $0.56 \pm 0.04$ & $0.65 \pm 0.07$  & $ 0.58 \pm 0.06$  \\
		\hline
	\end{tabular}
	\caption{Values of $r$ calculated in different rapidity regions for A0 and JH'2013 set 2
        gluon densities in a proton.}
	\label{tab:r_ratio}
	\end{table}%

\section{Numerical results} \indent

\begin{figure}
\begin{center}
{\includegraphics[width=.5\textwidth]{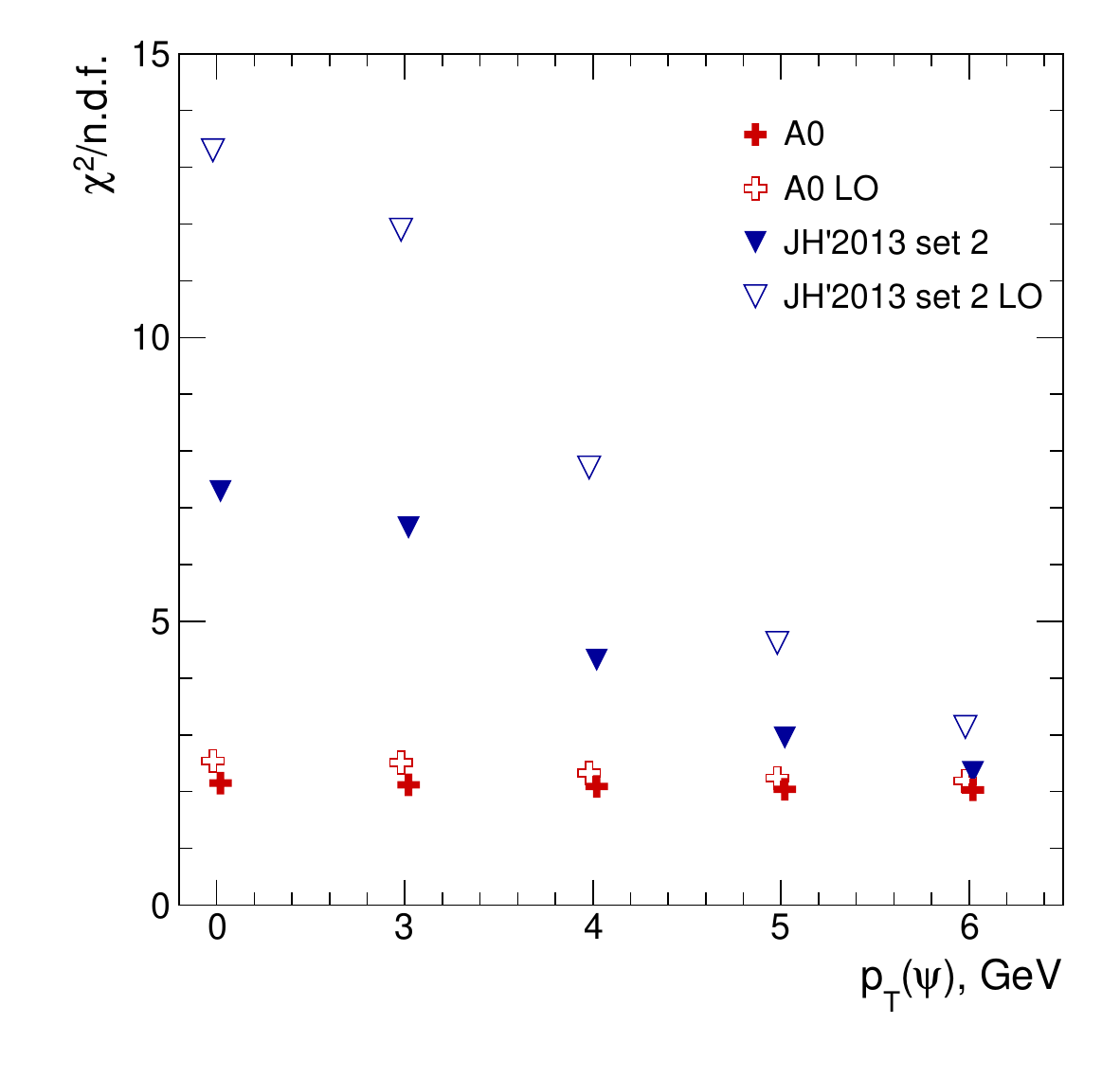}}\hfill
\caption{Lowest values of $\chi^2/{\rm n.d.f.}$ as function of minimal $\psi^{\prime}$  transverse momenta.}
\label{fig:Chi2_graph}
 \end{center}
\end{figure}

In this section, we present the results of our global fitting.
In Fig.~\ref{fig:Chi2_graph} we show the lowest values of $\chi^2/{\rm n.d.f.}$
obtained for different cutoffs $p_T^{\rm min}$ on the quarkonium transverse momentum.
For A0 gluon density, a good simultaneous description of all available LHC
data can be achieved without additional constraints on $p_{T}$.
Moreover, the results depend very weakly on the chosen $p_{T}^{\rm min}$.
An opposite situation is observed for JH'2013 set 2 gluon distribution,
where the larger $\chi^2$ values are obtained.
An essential part of $\chi^2$ comes from low $p_T < 6$ GeV at forward rapidities.
The large overestimation seen there (up to $3$ times) leaves no chance for a good
simultaneous description of the data. At the same time, excluding the problematic
area from the fitting yields results that are quite close to the A0 predictions.
The observed difference in the behavior of the $p_T$ spectra arises from the
dynamics of the gluon evolution in the considered TMD gluon densities
and could be used to distinguish between the latter (see below).

To highlight the importance of tree-level NLO corrections, we perform
	the same fitting procedure for the LO calculations. 
	The inclusion of NLO corrections leads to changing of the shape  and allows to significantly soften the discrepancy,
	especially for JH'2013 set 2 gluon (see Fig.~\ref{fig:Chi2_graph}).


\begin{table}[ht]
\centering
\small

\begin{tabular}[t]{lccc}
\hline
LDME &  \ A0  & \ JH'2013 set 2 & \ JH'2013 set 2 \footnotesize{($p_{T} > 6$)}  \\
\hline
\hline
LO+NLO$^{\dag}$  \\
\hline
$\langle\mathcal{O}_{l.c.}\rangle/$GeV$^{5}$  & $(6.06\pm0.07)\times 10^{-3}$ & $(6.17\pm0.06)\times 10^{-3}$ & $(9.95 \pm 0.09)\times 10^{-3}$  \\
$\langle\mathcal{O}[^3S_1^{[8]}]\rangle/$GeV$^{3}$  & $(10.2\pm0.2)\times 10^{-4}$ & $(10.7\pm0.1)\times 10^{-4}$  & $(7.46\pm0.13)\times 10^{-4}$  \\
\hline
LO \\
\hline
$\langle\mathcal{O}_{l.c.}\rangle/$GeV$^{5}$  & $(10.4\pm0.1)\times 10^{-3}$ & $(8.61\pm0.09)\times 10^{-3}$ & $(18.9 \pm 0.2)\times 10^{-3}$  \\
$\langle\mathcal{O}[^3S_1^{[8]}]\rangle/$GeV$^{3}$  & $(20.1\pm0.3)\times 10^{-4}$ & $(29.3\pm0.2)\times 10^{-4}$  & $(15.6\pm0.3)\times 10^{-4}$  \\
\hline
\end{tabular}
\caption{The values of LDME's for $\psi^\prime$ meson fitted with A0 and JH'2013 gluon densities in a proton.}
\label{tab:LDME}
\end{table}%

We wish to take a closer look at the comparison between the predictions based on
A0 and JH'2013 set 2 gluons without any $p_T$ cutoffs, and on JH'2013 set 2 gluons
at $p_{T} > 6$ GeV, where we have reasonable agreement with the data.
In Table~\ref{tab:LDME} we show the fitted LDME's values
for both LO and NLO$^{\dag}$ analyses.
The corresponding uncertainties are calculated as standard deviations from the
error matrix. 
The uncertainties are noticeably small because of large number of datapoints and
weak correlation between free parameters.

\begin{table}[ht]
\centering
\small
\caption{${\chi}^{2}/{\rm n.d.f.}$ for available data sets at LO+NLO$^{\dag}\, \vert$ LO.}
\begin{tabular}[t]{lcccc}
\hline
& \  n.d.f. \quad   & \ A0  & \ JH'2013 set 2 & \ \small{JH'2013 set 2} \footnotesize{($p_{T} > 6$)} \\
\hline
\hline
\footnotesize{ATLAS 5.02 TeV} \scriptsize{\cite{dataset_ATLAS_5}} & 15 \quad & 0.87 $\vert$ 0.96 &  1.69 $\vert$ 3.13 & 1.02 $\vert$ 1.52 \\
\footnotesize{ATLAS 7 TeV} \scriptsize{\cite{dataset_ATLAS7_8}}  & 168 \quad & 1.05 $\vert$ 1.12 & 2.77 $\vert$ 5.88 & 1.28 $\vert$ 1.65 \\
\footnotesize{ATLAS 8 TeV} \scriptsize{\cite{dataset_ATLAS7_8}}  & 172 \quad & 1.53 $\vert$ 2.25 & 2.15 $\vert$ 4.28 & 1.51 $\vert$ 1.70 \\
\footnotesize{ATLAS 13 TeV}  \scriptsize{\cite{dataset_ATLAS13}} & 84 \quad  & 5.54 $\vert$ 5.53 & 9.07 $\vert$ 16.4 & 4.89 $\vert$ 4.46\\
\footnotesize{CMS 5.02 TeV} \scriptsize{\cite{dataset_CMS_5}} & 10(8) \quad & 0.87 $\vert$ 0.59 & 4.26 $\vert$ 10.4 & 5.02 $\vert$ 14.5 \\
\footnotesize{CMS 7 TeV}  \scriptsize{\cite{dataset_CMS7}}   & 72 \quad  & 0.89 $\vert$ 1.39 & 3.69 $\vert$ 7.15 & 0.75 $\vert$ 0.88 \\
\footnotesize{CMS 13 TeV}  \scriptsize{\cite{dataset_CMS13}}  & 36 \quad  & 1.53 $\vert$ 2.06 & 9.20 $\vert$ 19.6 & 2.05 $\vert$ 3.41 \\
\footnotesize{LHCb 7 TeV}  \scriptsize{\cite{dataset_LHCb7_13}}  & 55(40) \quad & 3.27 $\vert$ 3.46 & 16.2 $\vert$ 28.1 & 4.66 $\vert$ 8.42 \\
\footnotesize{LHCb 13 TeV}  \scriptsize{\cite{dataset_LHCb7_13}}  & 79(59) \quad & 3.44 $\vert$ 4.37 & 25.1 $\vert$ 42.3 & 4.97 $\vert$ 8.49 \\
\footnotesize{ALICE 7 TeV}  \scriptsize{\cite{dataset_ALICE7}}  & 9(3) \quad & 1.09 $\vert$ 1.37 & 2.94 $\vert$ 6.21 & 1.82 $\vert$ 2.39 \\
\footnotesize{ALICE 8 TeV}  \scriptsize{\cite{dataset_ALICE8}}  & 9(3) \quad & 1.44 $\vert$ 1.62 & 1.99 $\vert$ 4.33 & 1.53 $\vert$ 1.78 \\
\footnotesize{ALICE 13 TeV}  \scriptsize{\cite{dataset_ALICE13}} & 12(6) \quad & 2.17 $\vert$ 2.93 & 4.28 $\vert$ 8.51 & 1.30 $\vert$ 1.71 \\
\hline
\hline
Summary      & 721(666) \quad & 2.15 $\vert$ 2.54 & 7.29 $\vert$ 13.3 & 2.34 $\vert$ 3.14 \\
\hline
\end{tabular}
\label{tab:Chi2_per_exp}
\end{table}%

Table \ref{tab:Chi2_per_exp} exhibits the $\chi^2/{\rm n.d.f.}$ values calculated
separately for each available data set (experiment + energy) with LDME's from the
Table~\ref{tab:LDME}. The values in parenthesis in the column "n.d.f." show the
number of datapoints remained after imposing cuts on transverse momentum $p_T>6$ GeV.
In the case of JH'2013 set 2 gluon, removing the $p_T$ constraints results in
a significant increase of $\chi^2$ for all LHCb and ALICE data sets and to an overall
letdown in the fit quality. This tendency is expectedly worse for LO only  (such problem was observed earlier \cite{Our-Charmonia-1}). It is worth noting the "ATLAS 13 TeV" data set, which
shows a significantly higher $\chi^2$ value compared to other experiments in the
central rapidity region. The one of reasons can be attributed to a significantly increased accuracy of the
data. We also note some discrepancy in the normalisation between the "ATLAS 13 TeV"
and "CMS 13 TeV" data. However, it is clearly seen that the overall improvement in fit quality is achieved for most datasets due to inclusion of tree-level NLO corrections.

\begin{figure}
\begin{center}
{\includegraphics[width=.49\textwidth]{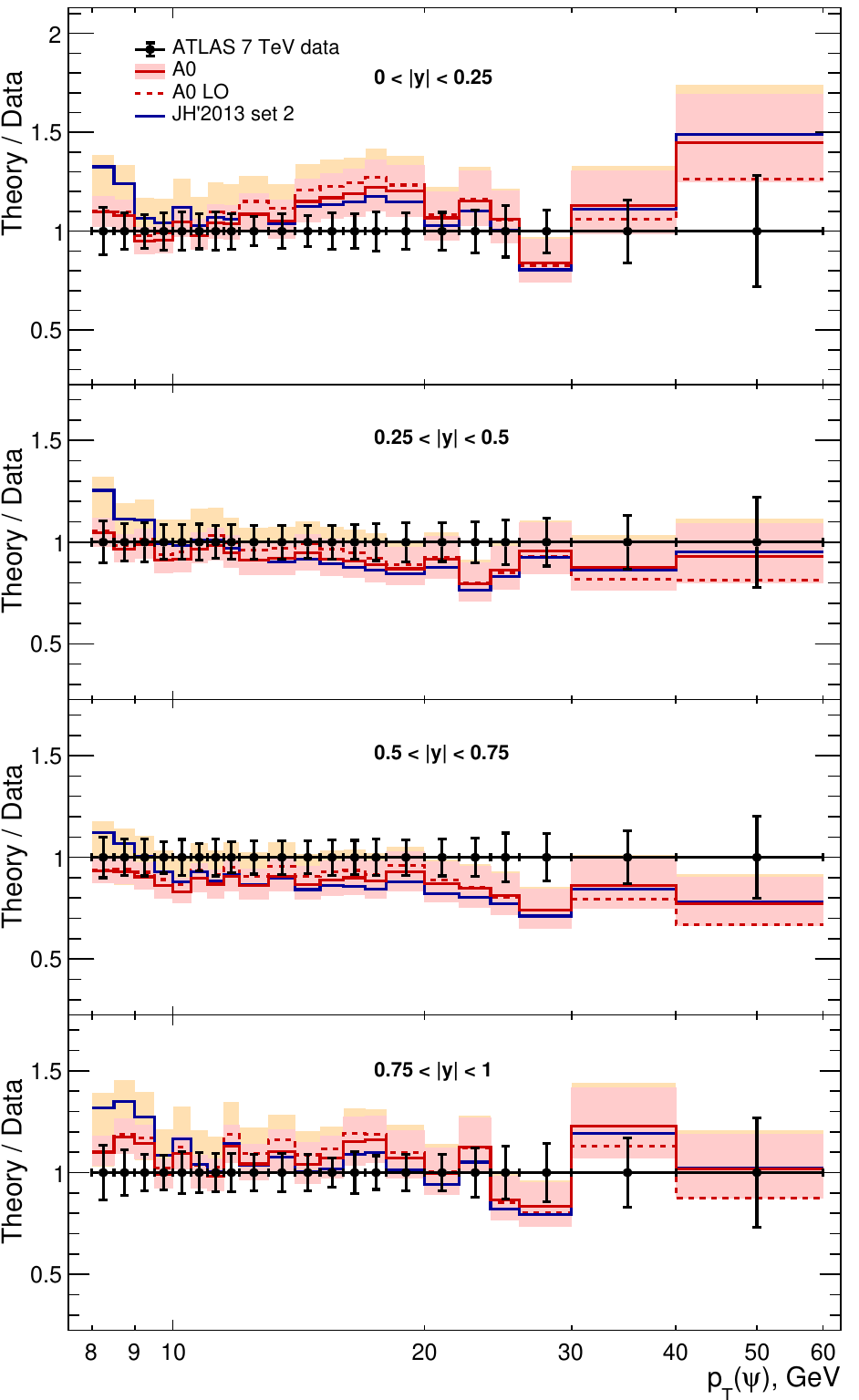}}\hfill
{\includegraphics[width=.49\textwidth]{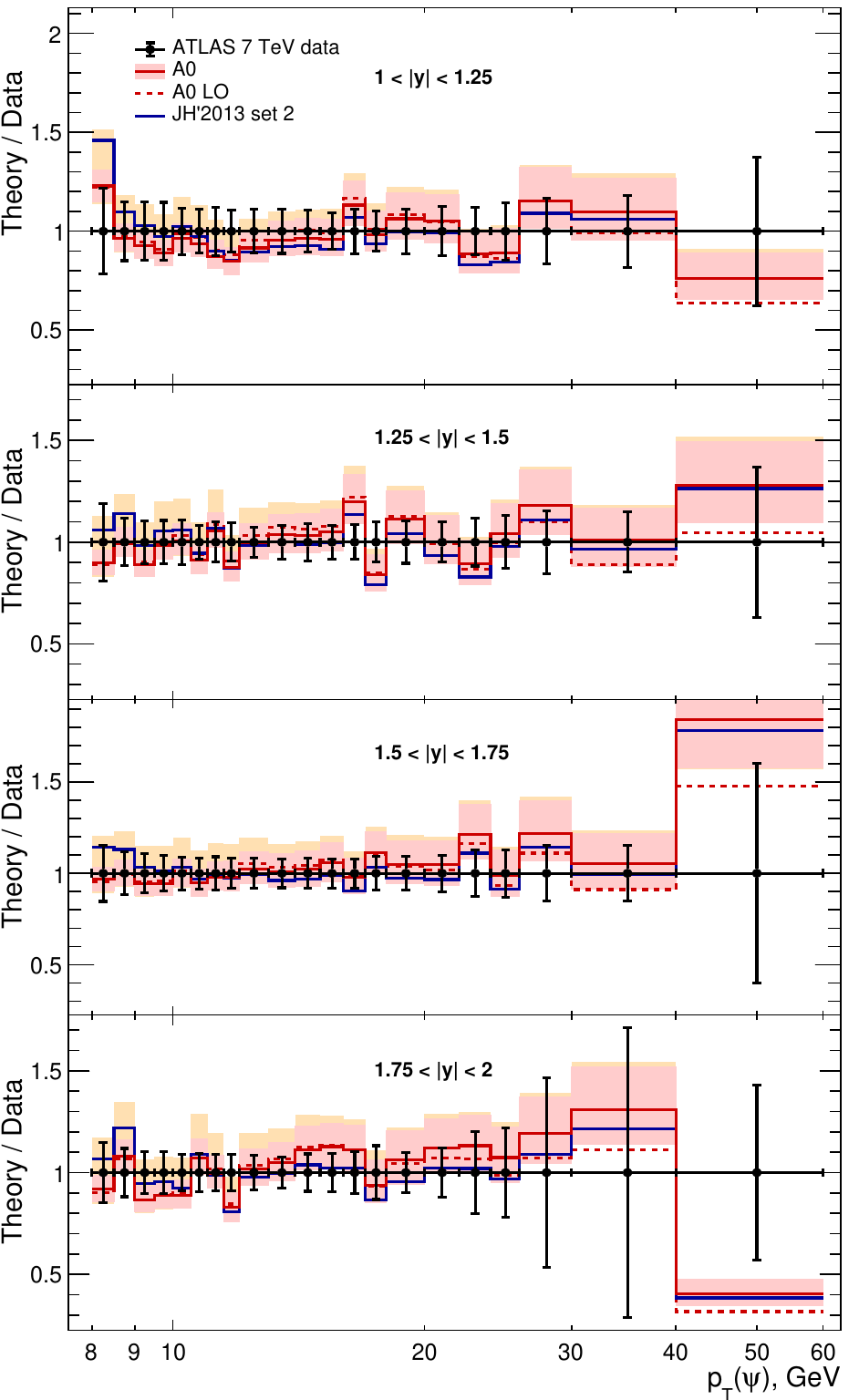}}\hfill
\caption{Ratios of differential cross sections for inclusive $\psi^{\prime}$ production in $pp$ collisions at $\sqrt{s} = 7$ TeV as functions of the $\psi^{\prime}$
transverse momentum at the central rapidity regions. The ATLAS data are taken from \cite{dataset_ATLAS7_8}.}
\label{fig:ATLAS_7TeV}
 \end{center}
\end{figure}

\begin{figure}
\begin{center}
{\includegraphics[width=.49\textwidth]{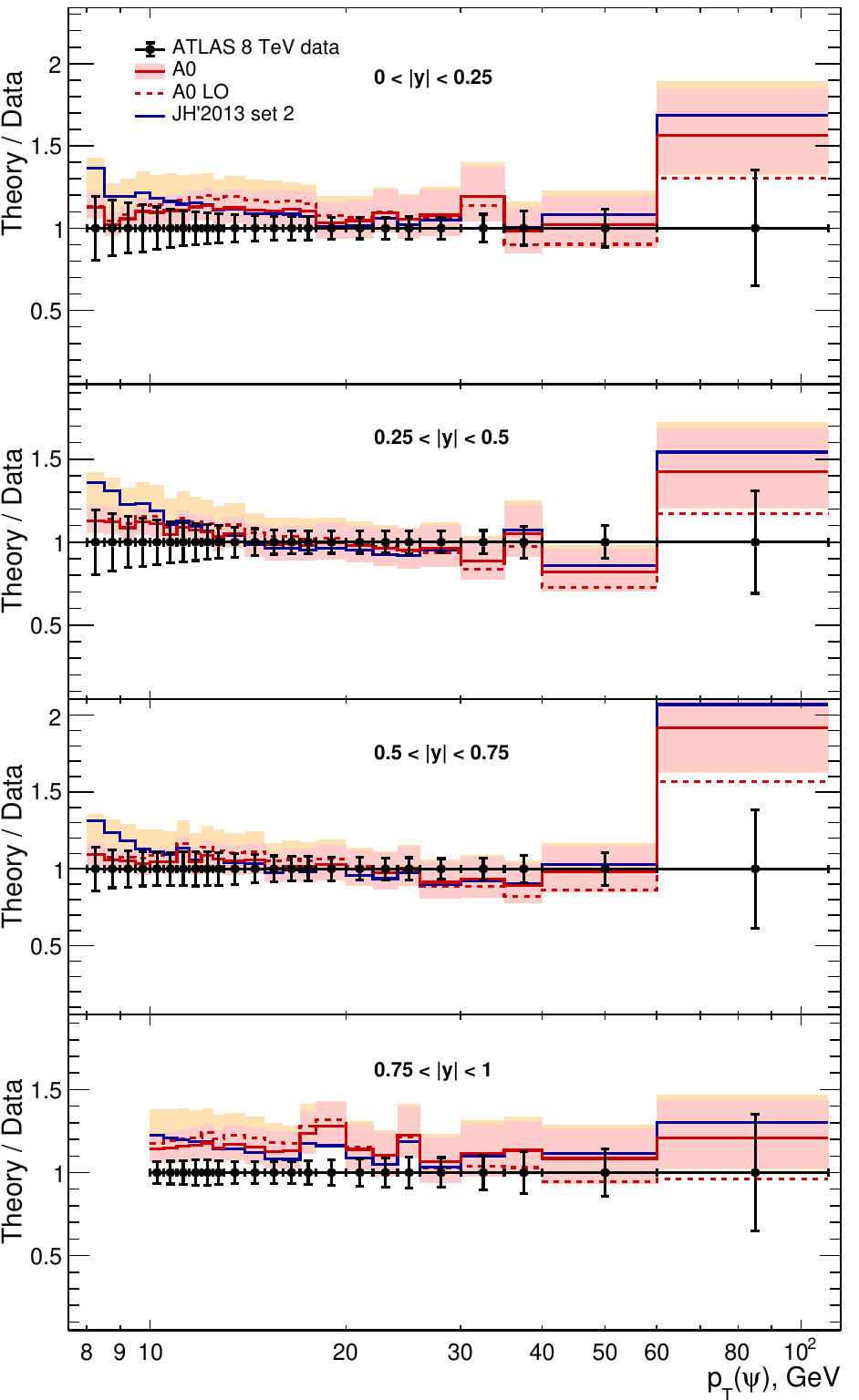}}\hfill
{\includegraphics[width=.49\textwidth]{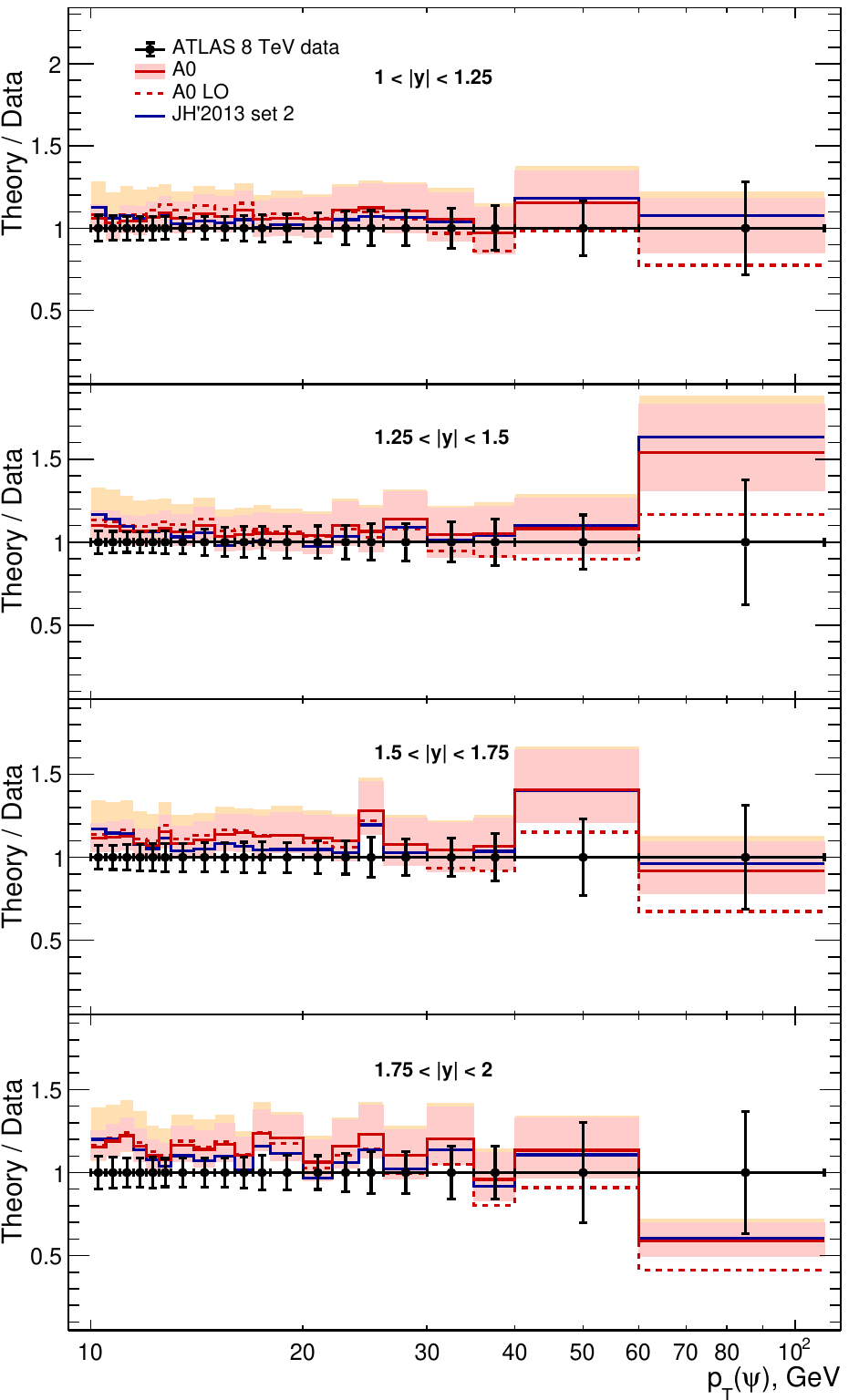}}\hfill
\caption{Ratios of differential cross sections for inclusive $\psi^{\prime}$ production in $pp$ collisions at $\sqrt{s} = 8$ TeV as functions of the $\psi^{\prime}$
transverse momentum at the central rapidity regions. The ATLAS data are taken from \cite{dataset_ATLAS7_8}.}
\label{fig:ATLAS_8TeV}
\end{center}
\end{figure}

\begin{figure}
\begin{center}
{\includegraphics[width=.49\textwidth]{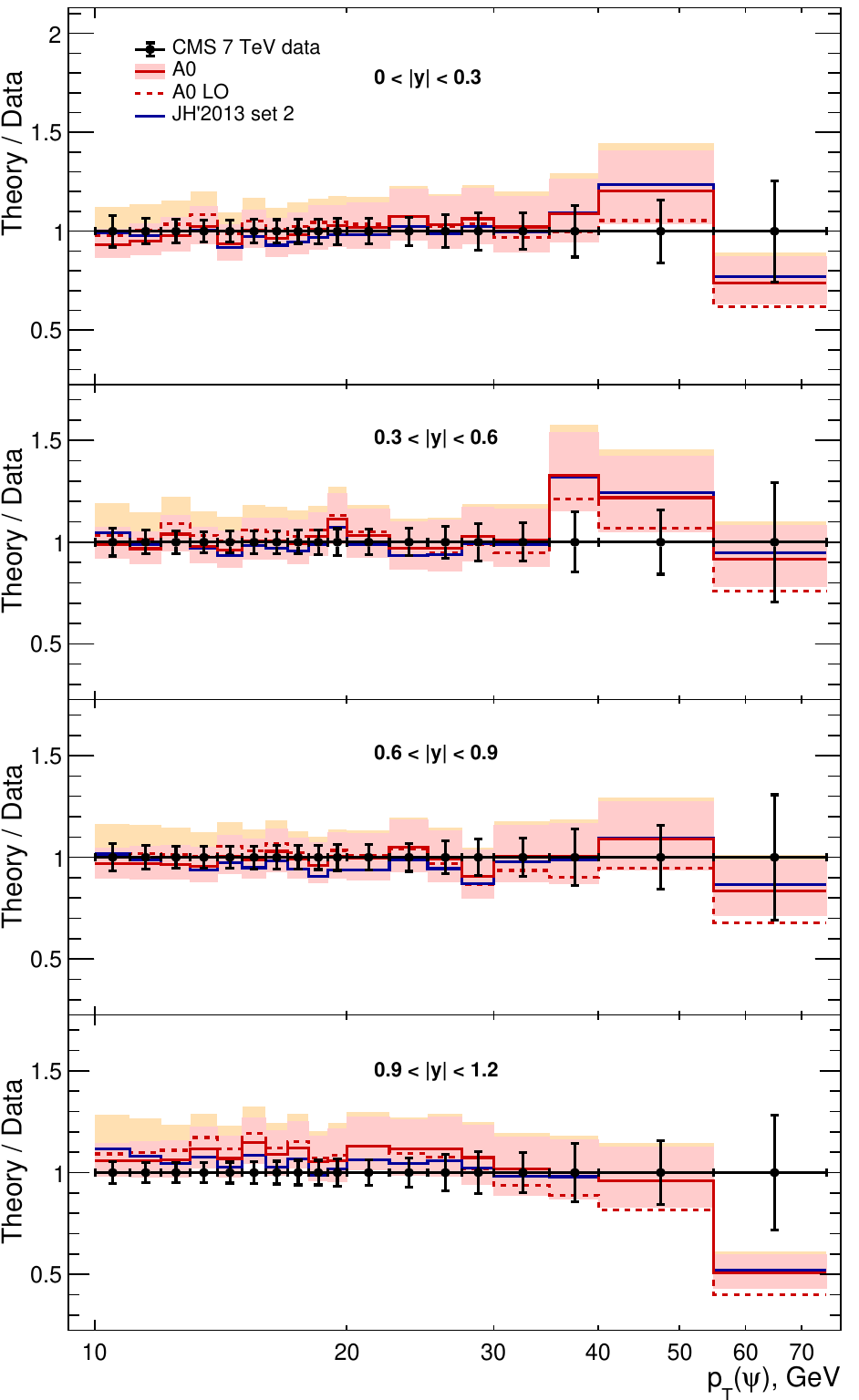}}\hfill
{\includegraphics[width=.49\textwidth]{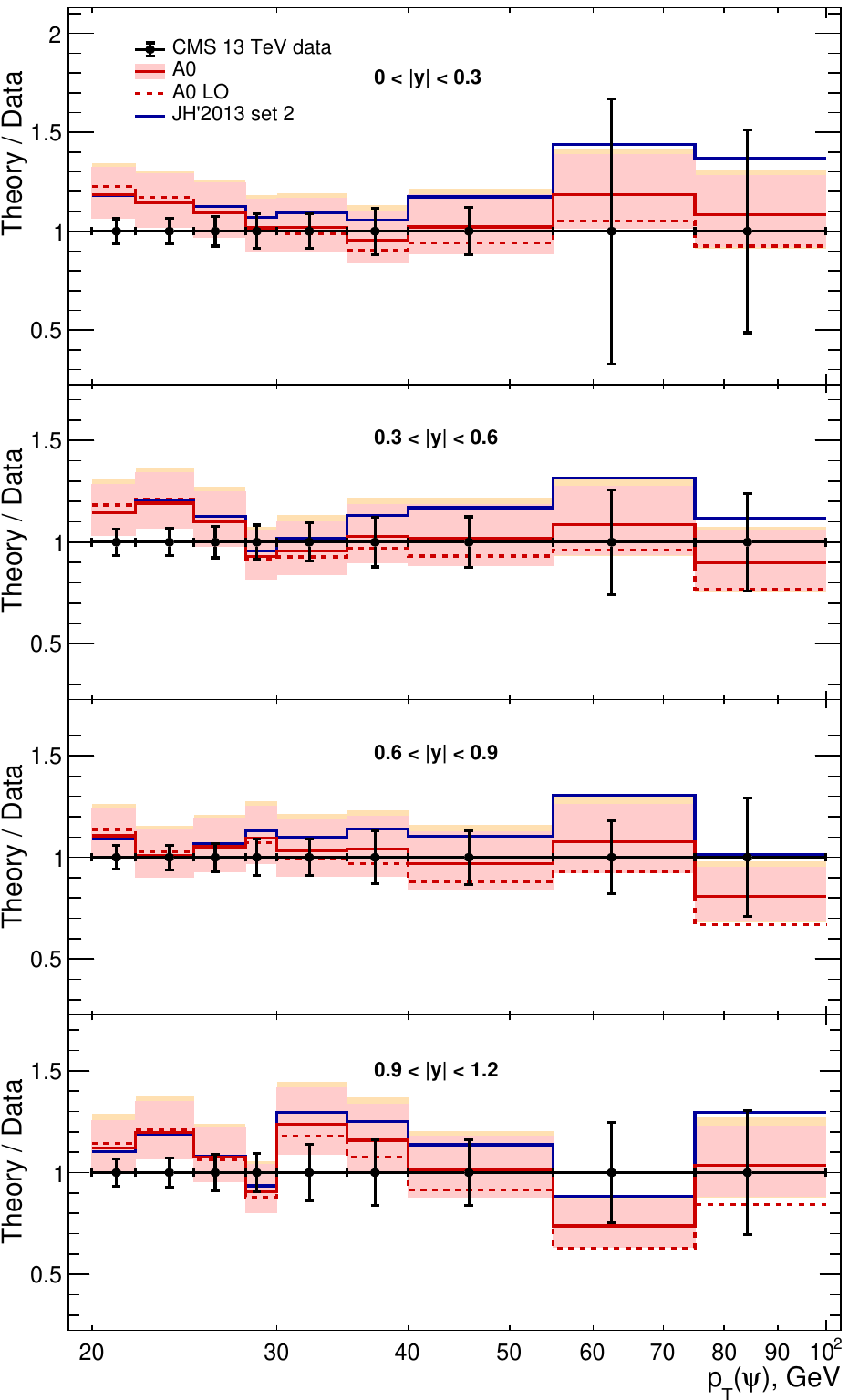}}\hfill
\caption{Ratios of differential cross sections for inclusive $\psi^{\prime}$ production in $pp$ collisions at $\sqrt{s} = 7$ TeV (left panel) and $\sqrt{s} = 13$ TeV as functions of the $\psi^{\prime}$
transverse momentum at the central rapidity regions. The CMS data are taken from \cite{dataset_CMS7} and \cite{dataset_CMS13}.}
\label{fig:CMS_7TeV}
\end{center}
\end{figure}

\begin{figure}
\begin{center}
{\includegraphics[width=.49\textwidth]{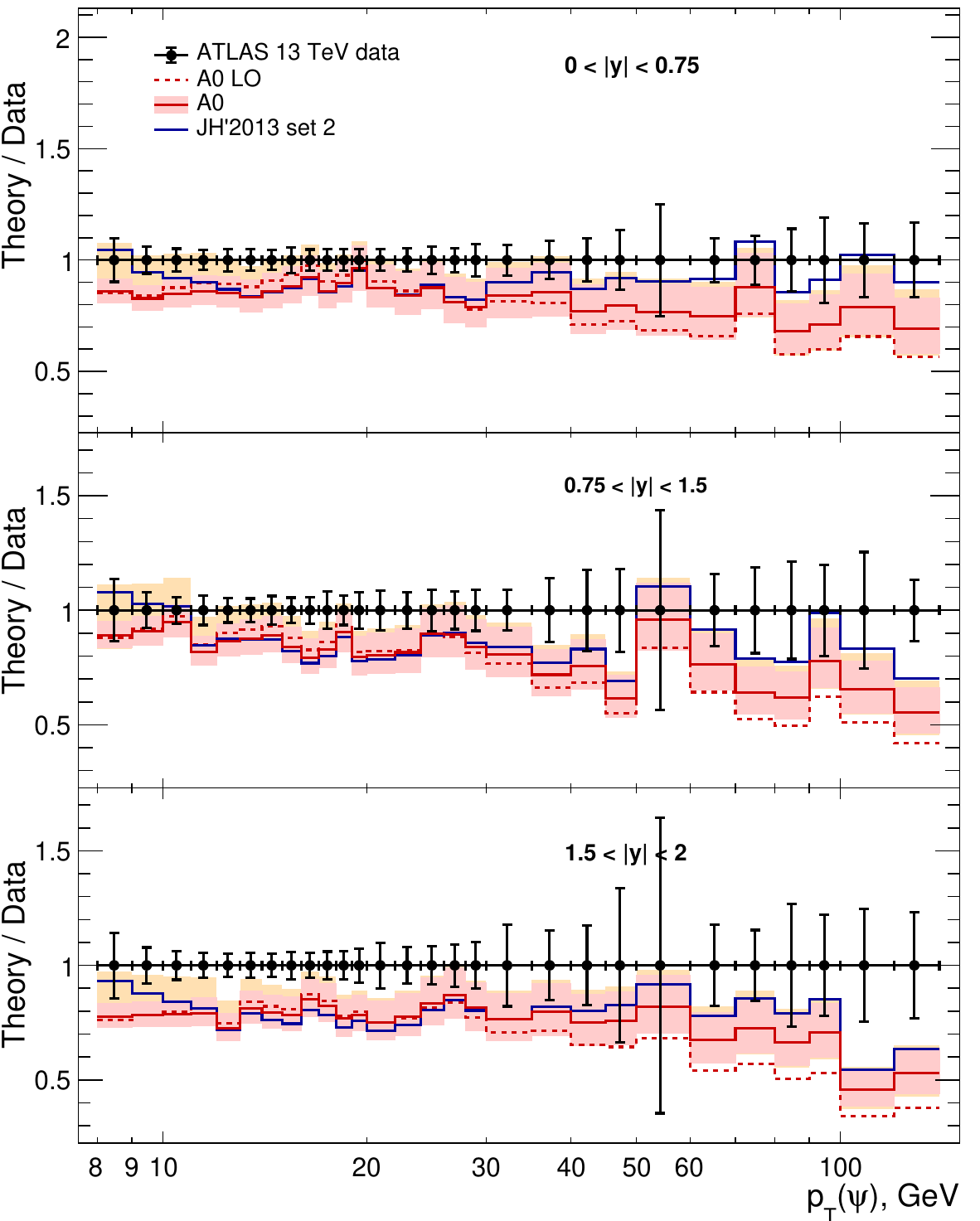}}\hfill
{\includegraphics[width=.49\textwidth]{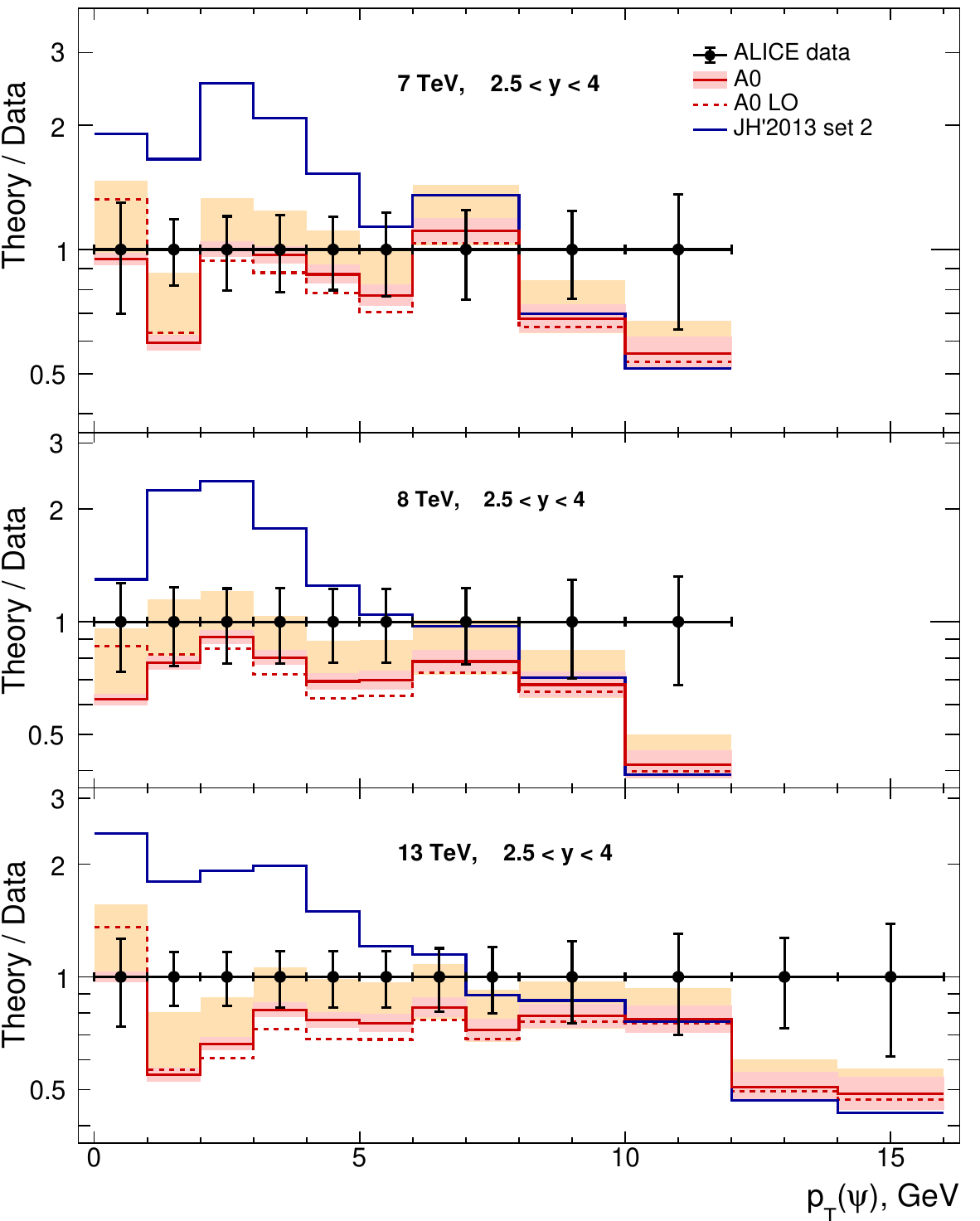}}\hfill
\caption{Ratios of differential cross sections for inclusive $\psi^{\prime}$ production in $pp$ collisions at $\sqrt{s} = 13$ TeV at the central rapidity regions (left panel) and $\sqrt{s} = 7,8, 13$ TeV at the forward rapidity regions (right panel) as functions of the $\psi^{\prime}$
transverse momentum. The ATLAS data are taken from \cite{dataset_ATLAS13} and ALICE \cite{dataset_ALICE7},\cite{dataset_ALICE8},\cite{dataset_ALICE13}.}
\label{fig:ATLAS_ALICE}
\end{center}
\end{figure}

\begin{figure}
\begin{center}
{\includegraphics[width=.49\textwidth]{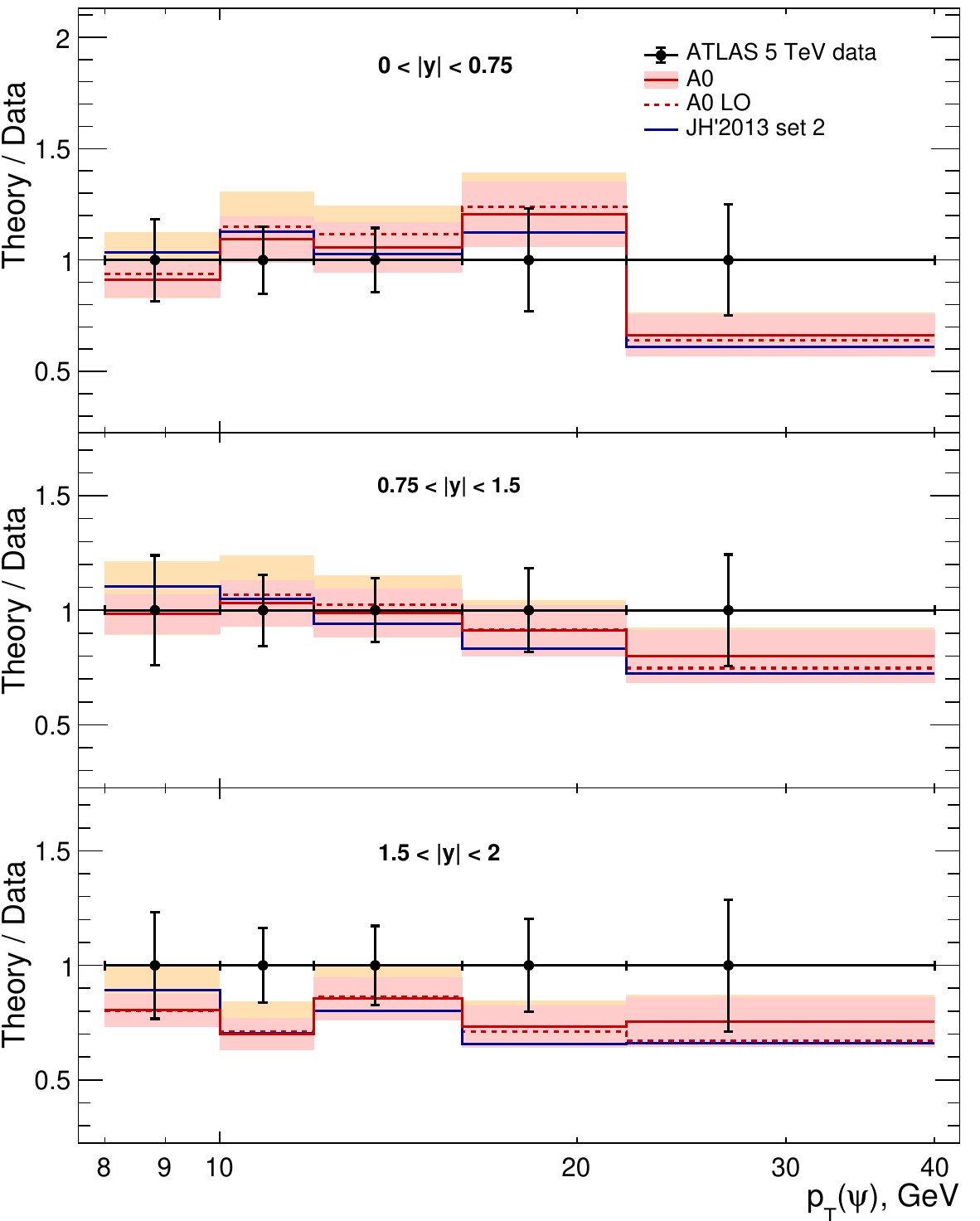}}\hfill
{\includegraphics[width=.49\textwidth]{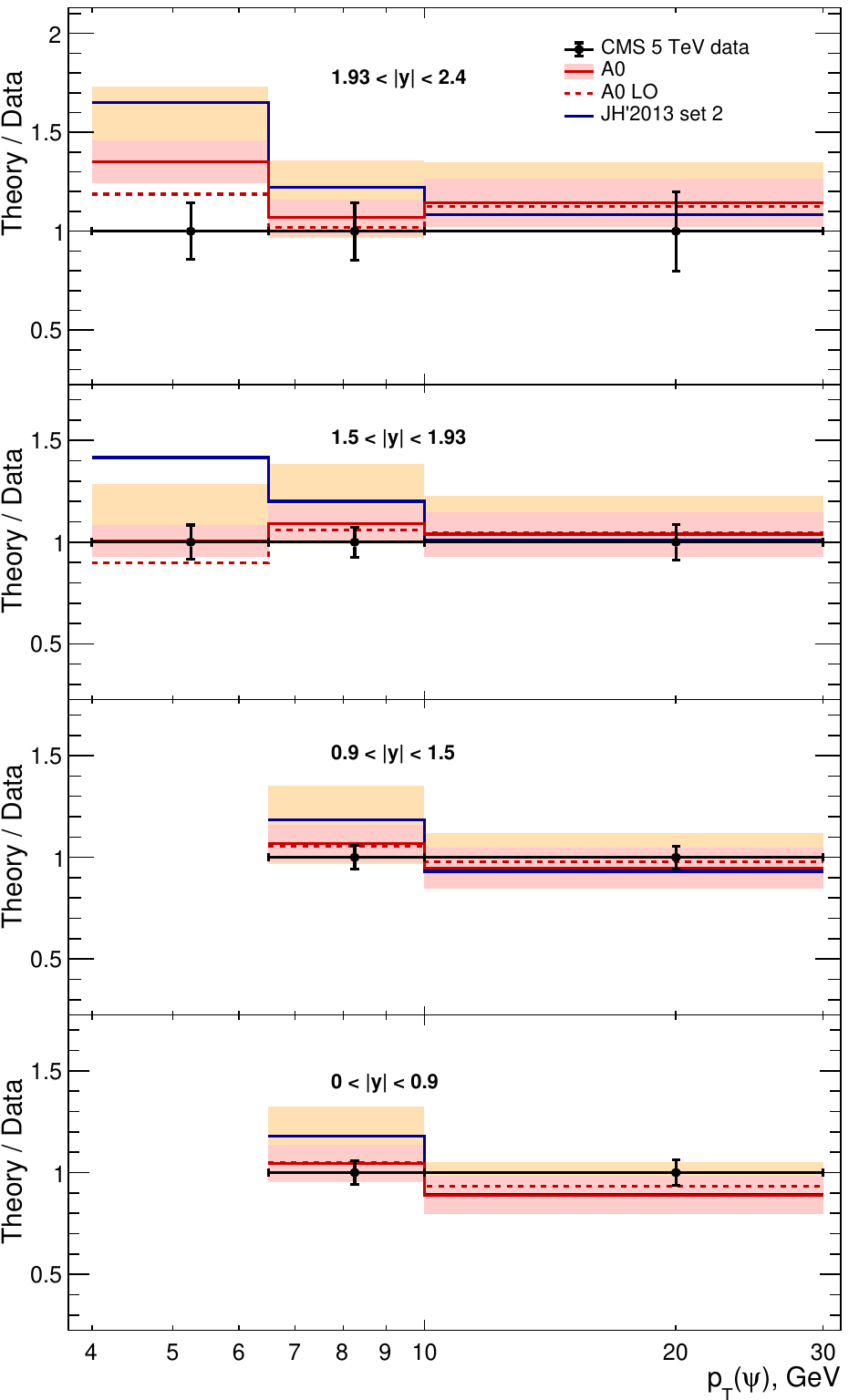}}\hfill
\caption{Ratios of differential cross sections for inclusive $\psi^{\prime}$ production in $pp$ collisions at $\sqrt{s} = 5.02$ TeV at the central rapidity regions as functions of the $\psi^{\prime}$ transverse momentum. The ATLAS data are taken from \cite{dataset_ATLAS13} and CMS \cite{dataset_ALICE7},\cite{dataset_ALICE8},\cite{dataset_ALICE13}.}
\label{fig:ATLAS_CMS_5}
\end{center}
\end{figure}

\begin{figure}
\begin{center}
{\includegraphics[width=.49\textwidth]{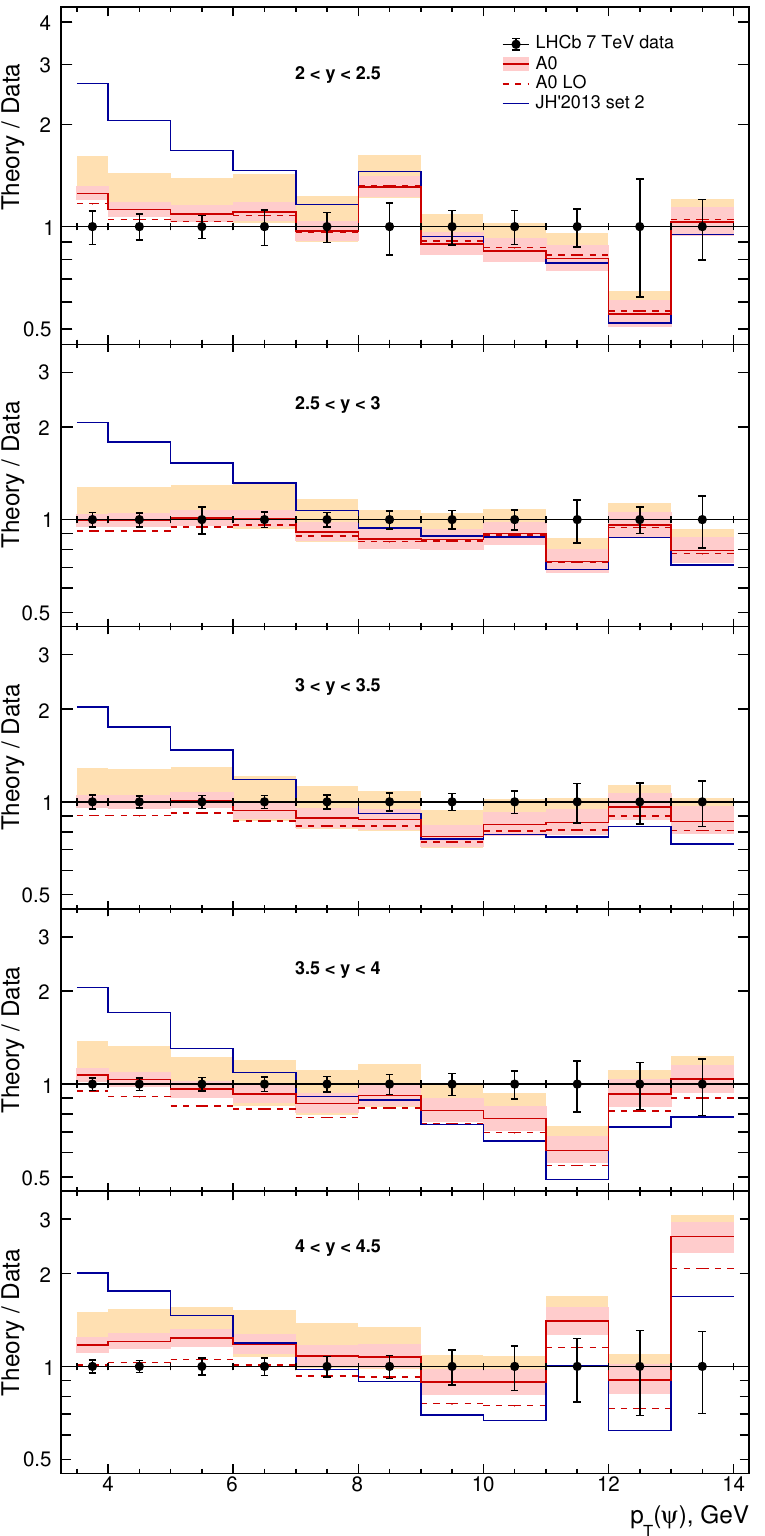}}\hfill
{\includegraphics[width=.49\textwidth]{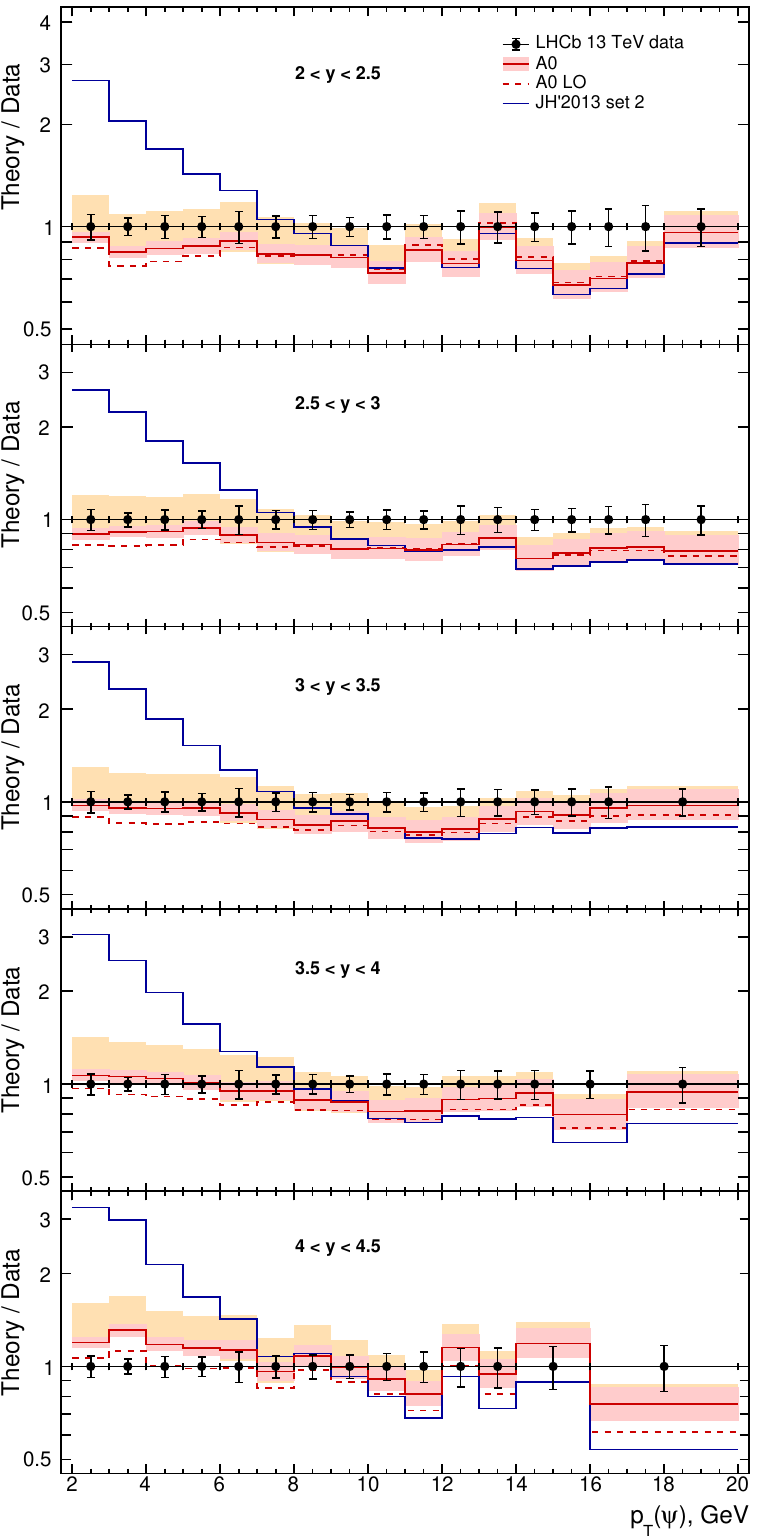}}\hfill
\caption{Ratios of differential cross sections for inclusive $\psi^{\prime}$ production in $pp$ collisions at $\sqrt{s} = 7$ TeV (left panel) and $\sqrt{s} = 13$ TeV (right panel) as functions of the $\psi^{\prime}$
transverse momentum in the forward rapidity region. The LHCb data are taken from \cite{dataset_LHCb7_13}.}
\label{fig:LHCb}
\end{center}
\end{figure}

The results of our calculations for the $\psi^{\prime}$ differential cross sections
are displayed in Figs.~\ref{fig:ATLAS_7TeV} --- \ref{fig:LHCb} for all data sets in
the form of "theory to data" ratios.
The results are only presented for A0 gluons without cutoffs (red lines) and
JH'2013 set 2 gluons with $p_T > 6$ GeV (blue lines).
The theoretical uncertainty bands (shaded areas) are shown for A0 gluon density.
The inner bands represent the reasonable variations in the soft gluon energy $E_g$ emitted at the
formation of bound states within the range $E_g=\Lambda_{\rm QCD}\pm 50$ MeV. The outer bands
include the renormalization scale uncertainties added in quadratures. As usual,
variations in the renormalization scale were accompanied with using A0$-$ and A0$+$
gluon densities for $\mu_R/2$ and $2\mu_R$ values, respectively (see\cite{A0}).
We find no dramatic difference between the results obtained with A0 and
JH'2013 set 2 gluon densities in the central and forward rapidity regions at
moderate and large $\psi^\prime$ transverse momenta, $p_{T} > 6$ GeV.
However, the region of relatively low $p_T$ shows a pronounced sensitivity
to the choice of TMD gluon distribution.
In addition, we show here the LO results for A0
	gluons to emphasize the difference in shape of $p_T$ spectra. The NLO tree-level corrections allow to visibly reduce the slope of spectra. It leads to better agreement with data in full available $p_T$ region despite the comparable $\chi^2$ for both LO and LO + NLO$^{\dag}$ cases.

One can conclude from Figs.~\ref{fig:ATLAS_7TeV} --- \ref{fig:LHCb} that
the modified LO + NLO$^{\dag}$ merging scheme developed to avoid double counting has
demonstrated its working power, and is thus validated for
future use\footnote{We leave the analysis of the $\psi^\prime$ polarization for the forthcoming dedicated study.}.
It will be implemented into the next version of the Monte-Carlo event generator \textsc{pegasus}\cite{PEGASUS}.

\section{Conclusion} \indent

We have considered inclusive $\psi^{\prime}$ production in proton-proton collisions
at the LHC energies in the $k_T$-factorization approach beyond the leading order
approximation.
Our work represents the first attempt to carry out a simultaneous description of
all available unpolarized data on $\psi'$ meson production in the entire kinematic region within high energy factorization approach.
The theoretical innovation consists in including the tree-level NLO$^\dag$ contributions
and in modifying the LO + NLO merging scheme for specific kinematic conditions
(forward rapidities at low $p_T$).

\vskip 0.3cm
\noindent
The results of our study can be summarized as follows.
\begin{itemize}
\item Adding the tree-level NLO contributions to hard partonic subprocess helped to
reach a reasonable agreement with the data in the whole kinematic range including the problematic region of low $p_T$, which has not yet been described satisfactorily in theoretical papers.
\item The modified LO + NLO$^{\dag}$ merging scheme developed to avoid double counting has
demonstrated its working power, and is thus validated for future use.
\item A significant part of theoretical uncertainties (mainly, regarding the overall
normalization) comes from the meson formation mechanism, namely, from the energy required for a soft gluon to overcome the confinement.
\item A comparison of the results obtained with different gluon densities shows no
significant difference between A0 and JH'2013 set 2 gluons in the region of moderate
and high $p_T$. The results obtained at small $p_T< 6$ GeV give clear preference
to the A0 gluons.
\end{itemize}

\section*{Acknowledgements} \indent

The study was conducted under the state assignment of Lomonosov State University.
This work has been supported also by the National Key R\&D Program
of China (Grant 2024YFE0109800 and 2024YFE0109802).

\bibliography{psi2S}

\end{document}